\documentstyle[epsfig]{aipproc}

\def\gtwid{\mathrel{\raise.3ex\hbox{$>$\kern-.75em\lower1ex\hbox{$\sim$}}}}
\def\ltwid{\mathrel{\raise.3ex\hbox{$<$\kern-.75em\lower1ex\hbox{$\sim$}}}}

\def\mnras{M.N.R.A.S.}
\def\apj{Ap.J.}
\def\apjl{Ap.J. Lett.}
\def\mn{M.N.R.A.S.}
\def\rmp{Revs. Mod. Phys.}

\begin{document}
 
\title{Magnetic Extraction of Spin Energy from a Black Hole}
    
\author{J.H. Krolik$^*$}

\footnote{$^*$Department of Physics and Astronomy, Johns Hopkins
University, 3300 N. Charles St., Baltimore, MD 21218; jhk@pha.jhu.edu}
 
%\maketitle

\begin{abstract}

    Numerous variations have been proposed on the original suggestion by
Blandford and Znajek that magnetic fields could be used to extract rotational
energy from black holes.  A new categorization of these variations is
proposed so that they may be considered in a systematic way.  ``Black
hole spindown" is defined more precisely, distinguishing decrease in
the spin parameter $a/M$ from decreases in angular momentum $a$
and rotational kinetic
energy, $M - M_i$. Several key physical questions are raised: Can the
``stretched horizon" of a black hole communicate with the outside world? 
Do accretion disks bring any net magnetic flux to the black holes at
their centers?   Is the magnetic field adjacent to a black hole force-free
everywhere?

\end{abstract}

\section{Background}

     Adding angular momentum to a black hole while keeping its mass fixed
decreases its surface area:
\begin{equation}
A = 8\pi M^2 \left[ 1 + \sqrt{1 - a_*^2}\right], 
\end{equation}
where, as usual, $M$ is the black hole mass, $a_* \equiv a/M$
is its dimensionless spin parameter with $a$ the specific angular
momentum, and all quantities are in relativistic units
(i.e., $G = c = 1$, so that, for example, the unit of length is $r_g
= GM/c^2$).  Hawking and Ellis \cite{jhk:he73}
showed that $A$ cannot decrease; consequently, no matter what happens, a
black hole of surface area $A$ cannot ever have a mass less
than $(A/16\pi)^{1/2}$, its ``irreducible mass" $M_{i}$.  However, if $a_* > 0$,
$M > M_i$.  Therefore, reducing $a_*$ can make $M$ smaller; that is, braking
a black hole's spin can make it lose energy.  In principle, this energy can
be delivered in usable form to the outside world.  If $a_* = 1$, the energy
potentially tappable is $(1 - 1/\sqrt{2})M \simeq 0.293 M$.  Comparing this
quantity to the amount of energy released in the course of accretion
(generally estimated as $\sim 0.1M$), we see that there is potentially as
much energy stored in black hole spin as can be released in ordinary accretion.

      The first proposal of a specific mechanism to extract this energy
was due to Penrose \cite{jhk:p69}.  His scheme made use of the fact that particles
inside the ergosphere can achieve {\it negative} total energy
(i.e., including their rest-mass energy) if their orbital frequencies are
small enough ($\Omega < a^{-1}(1 - r/2M)$).  However, Bardeen et al.
\cite{jhk:b72} pointed out that it would be
very difficult to accomplish this task by particle-particle events because there
is an extremely large velocity difference between positive and negative energy
orbits.

      Since Penrose's initial suggestion, a number of other ideas have been
proposed.  The unifying theme of all these proposals is to couple the black
hole's spin to external matter by magnetic forces.  In some sense, they are
non-local realizations of the Penrose process that make use of electromagnetic
effects to convey negative energy into the hole, whether by injecting it
with negative energy wave modes or negative energy particles.

      Because the first such scheme was proposed by
Blandford \& Znajek \cite{jhk:bz77},
there has been a tendency in the literature to refer to them all generically
as the ``Blandford-Znajek" mechanism.  Although all of them share the
fundamental idea of magnetic couplings, and so in that sense are different
versions of the same basic idea, there are now enough variations on the theme
to make a clearer categorization useful.  It is the object of this paper to
present such a categorization, along with some comments about these variations'
relative standing.

\section{The Rate of Rotational Energy Change}

    Before embarking on this discussion it is
worthwhile to pause a moment to discuss several points of principle  in order
to clarify our language.   First of all, let us list the ways by which
black holes can gain or lose angular momentum (we will omit mechanisms
involving singular events such as the initial creation of the black hole or a
merger with a comparable mass object).  One way is through the action of its
own gravitational field: matter orbiting with an angular momentum vector
oriented obliquely to the angular momentum of a black hole feels a
torque.  Here we will ignore this effect, assuming that any nearby matter
orbits in the plane normal to the black hole spin.
A second way is through matter crossing the event horizon bearing
angular momentum.  The third and final mechanism is through
capture of photons with orbital angular momentum.

    Next, let us consider what is required in order to identify the source
of energy for a given event.  It is occasionally assumed (e.g., in \cite{jhk:lop98})
that all energy lost from stably-orbiting matter in an accretion disk is
drawn from gravitational potential energy lost by
accreting matter, with none coming from the spin of the black hole.  However,
this distinction---made locally---can be problematic because the shape of the
gravitational potential depends on the black hole spin.  To take an extreme
example, when $a_* = a/M > 0.943$, the marginally stable orbit falls inside the
ergosphere; in that case, part of the orbital kinetic
energy of the disk material is due to the rotation of the black hole.  If
there are other ways of coupling the rotation of the black hole to the
disk (see below), the identity of the source of energy at any particular
location can become even fuzzier.

     Although the local energy source can be ill-defined, it is always
possible to make this distinction globally.  All that is necessary is to
compute the fluxes of angular momentum and energy across the black
hole's event horizon.

   However, even for this distinction one must be careful about definitions.
For example, if matter accretes with exactly the specific angular momentum
$\hat L_{ms}M$ of the marginally stable orbit and the angular momentum of
captured photons is neglected, the spin parameter $a_*$ always increases toward unity; when captured radiation is included in the accounting, $a_*$ reaches equilibrium at $\simeq 0.998$ \cite{jhk:t74}.  If the
accreting matter arrives at the hole with a fraction $1 - {\cal L}_{ms}$ of
$\hat L_{ms}M$, we may regard the spin as being tapped because the
hole's rotational energy is less than it would have been if only the usual
amount of energy (i.e. the binding energy at the marginally stable orbit) had
been released.

      One might also choose to impose the stronger condition
that the hole is being ``spun down" if $a_*$ actually decreases.  Dividing
the angular momentum fluxes into those depending on accretion of rest-mass
and those depending on photon capture (photon capture also
includes capture of electromagnetic waves), we find that the rate
of change of $a_*$ is
\begin{equation}
{d a_* \over dt} = 
{\dot M_o \over M} \left[ \left(1 - {\cal L}_{ms}\right)\hat L_{ms}
     - 2 a_*(1-\eta)\right] + {\dot M_{\gamma} \over M}
     \left[ {\cal J}_{\gamma} - 2a_*\right],
\end{equation}
where $\dot M_o$ is the rate of rest-mass accretion,
$\eta$ is the fraction of rest-mass energy lost before matter
arrives at the black hole, $\dot M_\gamma$ is the rate at which the
black hole mass changes due to photon accretion, and ${\cal J}_\gamma M$
is the mean angular momentum per photon.  Note that $\dot M_\gamma$
depends on the energy of photons as measured at infinity; that can be
quite different from the locally-measured photon energy.  If photons
are unimportant, whether $a_*$ increases or decreases depends on the
balance between the angular momentum brought into the black hole
$(1 - {\cal L}_{ms})\hat L_{ms}$ and the angular momentum required to
maintain the same ratio of spin to mass, $2a_* (1 - \eta)$.

    It is apparent from equation 2 that $a_*$ can fall even while the
angular momentum $a$ increases: the rate of change of $a$ is the same
as for $a_*$ but without the two terms $\propto a_*$.  Moreover, as equation~2
also shows, $a_*$ might remain constant even while $M$ increases.
If that is so, the total rotational energy
still increases.  This fact suggests a still stronger definition--that the
actual rotational energy $M - M_i$ decreases.  Writing this criterion
in terms of its component quantities, we have
\begin{eqnarray}
{d(M - M_i) \over dt} = &\dot M_o\left\{
  \left(1 - \eta\right)\left[1 - m_i -
     {a_*^2 \over 2 m_i(1 - a_*^2)^{1/2}}\right]
  + {a_* \hat L_{ms} (1 - {\cal L}_{ms}) \over 
                  4 m_i(1 - a_*^2)^{1/2}}\right\} \nonumber \\
   &+ \dot M_\gamma\left[1 - m_i - {a_*^2 \over 2 m_i(1 - a_*^2)^{1/2}}
         +{a_* {\cal J}_\gamma \over 4 m_i (1 - a_*^2)^{1/2}}\right],
\end{eqnarray}
where $m_i \equiv M_i/M$.

   Thus, we see that the vague term ``spindown" can connote any of several
distinguishable results: diminishing $a_*$, $a$, or $M - M_i$.
Because rotation contributes only a small amount to the
total energy of the black hole when $a_*$ is small, the different criteria
are very similar for $a_* \ltwid 0.8$;
for larger values of $a_*$, the difference becomes more significant.  When
considering stored-energy reservoirs, the last definition is the most
precise.

\section{The Original Blandford-Znajek Mechanism}

   As remarked above, the first plausible mechanism for removing
black hole spin energy was proposed by
Blandford \& Znajek \cite{jhk:bz77}.  They
imagined that as a black hole accretes, it would inevitably trap some net
magnetic flux due to accreted field lines with connections to infinity.
There would then be an approximately time-steady magnetic field configuration
with field lines embedded in the hole's event horizon, even while their
far ends close at very large distance from the black hole.  The enforced
rotation of space-time due to the black hole spin would then
drive an MHD wind.  In their initial formulation, the field structure was
supposed to be force-free everywhere, i.e. there was so little plasma attached
to these field lines that $B^2/4\pi \gg \rho c^2$ everywhere.
Phinney \cite{jhk:p83}
extended this picture to include a small, but finite, plasma rest-mass
density so that MHD wave group speeds would remain (slightly) less than
$c$.  Even in this modification, there is so little inertia outside the
event horizon of the black hole that the orbital motion of nearby plasma
has little effect on the rotation rate of magnetic field lines; only
the distant region where the energy is delivered is considered to have
any significant inertia.

      Similarly, because the accretion rate is taken to be negligible,
the change in the black hole's rotational energy is due solely to
the term proportional to $\dot M_\gamma$ in equation 3; i.e., negative
angular momentum and energy are brought to the black hole by zero rest-mass
electromagnetic waves, not by ordinary matter.

    These processes may also be visualized in terms of their
associated electric fields.  Suppose that the magnetic field is
stationary with respect to a distant observer.  Matter just outside the
black hole is compelled to rotate with the hole, so in the matter's
frame there is
an electric field perpendicular to both the field and the rotational
velocity whose magnitude is proportional to the velocity mismatch
between the field frame and the matter frame.
This field drives a current from one field-line to another
through the resistance of the horizon; the current then flows out along
the new field-line until somewhere far from the black hole
it crosses back to its original field-line and closes the circuit.
When the resistance at infinity $R_l$ is comparable to the
resistance in the horizon $R_h$, energy is dissipated in the distant
``resistor" at a rate $\sim c r_g^2 B_h^2$, where $B_h$ is the magnitude of the
field at the event horizon.  This impedance
matching is equivalent to the condition that all the work done by the horizon
forcing the field to rotate reaches infinity.

      In the usual formulation, the magnetic field is taken to be
time-independent and force-free everywhere
between the black hole and a distant, but localized, ``load".  A corollary
to the assumption that the load is localized is that ideal MHD
and flux-freezing apply everywhere between the black hole and the load.
The force-free assumption then assures that wherever the field and the
plasma move in lock-step the field controls the rotation rate.
Given those assumptions, matching the current through the horizon to
the current through the load determines the rotation rate of the flux-lines
linked by the current loop:
\begin{equation}
\Omega_F = {\Omega_h R_l + \Omega_l R_h \over R_h + R_l},
\end{equation}
where $\Omega_h$ is the rotation rate of the black hole and $\Omega_l$
the rotation rate of the load \cite{jhk:b90}.  The load's
rotation rate is determined by a balance between whatever forces
(gravitational, magnetic, $\ldots$) act upon it.

    When thinking in terms of the equivalent circuit, it is important to
distinguish true dissipative resistance from effective resistance (this
distinction is closely related to the distinction between ordinary and
radiation resistance).  For example, if the field-lines at infinity pass
through plasma with high conductivity, the major part of the energy
transmitted to infinity can be in the form of bulk work; i.e., the
distant matter can be accelerated coherently.  By contrast, ordinary
resistance produces heat.  The impedance matching
referred to earlier refers to the total load at
infinity, not solely the dissipative part.

    Because $B_h$ depends on the history of past accretion onto the black hole,
specifically the total amount of net magnetic flux accreted, it is unclear how
large $B_h$ should be.  Because there is resistance in the horizon, the
field must be supported by currents somewhere in the vicinity of the black
hole.  Rees et al. \cite{jhk:r82} suggested that they might be located in a ``fossil"
accretion disk so that they might not be any farther from the horizon
than a distance $\sim r_g$.  Rees et al. then argued that the field
could not be any larger than $\sim (r_{ms}/r_g)^2$ times the pressure in
the fossil disk, or there would be no dynamical equilibrium.
Another way to set the scale is to suppose that there is
a small amount of accretion, and estimate $B_h$ as comparable to the field
strength in the nearby accretion disk; if, as seems likely from recent
numerical simulations \cite{jhk:b96,jhk:s96}, 
$B^2 < 8\pi p$, this estimate yields a result somewhat smaller than
the previous one \cite{jhk:ga97}.  Combining the more conservative
estimate with the assumption of good impedance matching,
the expected luminosity is
\begin{equation}
L_{BZ} \sim \dot M_o c^2 (v_{orb,ms}/c) (r_{ms}/h_{ms}) (r_g/r_{ms})^2 ,
\end{equation}
where everything with subscript $ms$ is evaluated at the marginally stable
orbit.

    There are several issues regarding this mechanism that remain incompletely
understood.  Punsly \& Coroniti \cite{jhk:pc90a} raised the question of whether
the black hole event horizon and plasma far from the black hole could be in
causal contact.  They argued that, although the plasma density is taken to
be negligible in the usual formulation of the Blandford-Znajek mechanism,
it cannot be literally zero.  There would then be some accretion onto
the black hole, and the fast magnetosonic speed would be (slightly) less
than $c$.  If so, there would inevitably be a surface surrounding the
event horizon within which the inward velocity of the plasma would be
greater than the fast magnetosonic speed, and no signal carrying energy or information could propagate outward across that surface. 
Punsly \cite{jhk:p96} further argued that, because of severe
gravitational redshifting, electromagnetic waves (in this case, the fast
magnetosonic mode) can carry only tiny amounts of energy away from the event
horizon.  The significance of these criticisms is still unclear.   

     Further questions may also be raised about this form of
the Blandford-Znajek scenario:  Must the black hole necessarily accumulate
a net flux?  Perhaps all field lines suffer enough reconnection in the
course of accretion that only closed loops are brought to the black hole.  If
this is the case, the field loops would close near the black hole, not far
away.  Would the mechanism still work if the plasma is not
magnetically-dominated everywhere between the black hole and the load?
Field lines passing through the accretion disk, for example, will surely have
substantial inertia attached.  Similarly, the load region (presumably
near the Alfvenic surface of the wind where $B^2/4\pi \sim \rho v^2$)
might not be too far from the black hole even if the field lines close at
much greater distance.  One immediate consequence of
non-force-free behavior would be to change equation~4 to involve
a mean value of $\Omega_l$ weighted inversely by the local resistance
between the magnetic flux surfaces.  Two other consequences would
be more serious, however.  Significant plasma inertia would cause
the magnetosonic speed to fall well below $c$, so that the inner magnetosonic
surface would likely move well outside the event horizon.  Plasma inertia
would also slow the approach to stationarity and possibly disrupt it altogether.  In that case, the whole time-steady picture---a fixed
magnetosonic surface, well-defined field rotation rates, a steady-state
lumped-parameter electrical circuit analog---would be undercut.

     Differing answers to these questions may be regarded as the basis for
the variations on this scheme that have been suggested.  Alternatively, they
may be used as the structure of a classification scheme.  We will organize
the remainder of this paper along those lines, dividing schemes according
to their answers to three questions:

\noindent $\bullet$ Does the innermost part of the magnetic field run
through the event horizon or through plasma in the ergosphere?

\noindent $\bullet$ Do the field loops close nearby (e.g., in the disk)
or far away?

\noindent $\bullet$ Is there anywhere in the system other than the horizon
and a localized load where plasma inertia matters? 

    Labelled by this scheme, the original Blandford-Znajek mechanism
is one in which the field lines run through the event horizon and out to
infinity, and the field is force-free everywhere except in the event
horizon and the load.

\section{Field Lines Anchored Outside the Horizon, Closed at Infinity, and
Force-Free}

     The first alternative was invented by Punsly and Coroniti
\cite{jhk:pc90a,jhk:pc90b} and elaborated by Punsly (1996).
It has also been recently discussed by
Li \cite{jhk:l00b}.  They suggested that a better means
to extract the rotational energy of the black hole might be a
magneto-centrifugal wind anchored in plasma orbiting in the ergosphere,
but well away from the event horizon (see also \cite{jhk:o92} for a similar
idea).  This device would finesse the possible causality problem of field
lines tied to the event horizon, but would be otherwise quite similar to
the original Blandford-Znajek idea: the field lines extend to infinity,
the structure is supposed at least roughly time-steady, and the plasma
inertia is required to be low enough as to satisfy the condition of
magnetic domination everywhere but possibly in the distant load.

     Preliminary numerical simulations of a version of this idea exist
\cite{jhk:ksk98}.  In these simulations, the initial magnetic field
was taken to have uniform intensity and to be directed parallel to the
rotation axis.  Although the Alfven speed is relatively small in the
equatorial plane, it is $\simeq c$ everywhere outside the disk.  As
a result, magnetic braking of orbiting material is very strong, leading
to dramatic infall, shocks, and a (transient?) jet that is
pressure-driven along the axis but magnetically-propelled at larger radius.
Just as for the classic Blandford-Znajek mechanism,
the characteristic luminosity scale is $\sim |\vec B|^2 r_g^2 c$, but the
coefficient could be considerably less than unity.
Based on this point of view, Meier et al. \cite{jhk:m97} and
Meier \cite{jhk:m99} have also sugested that the relative strength of the magnetic
field at the base of the jet could explain the morphological contrast
between FR1 and FR2 radio galaxies.

\section{Field Lines Anchored in the Horizon, Closed in the Disk, and
Force-Free}

    In the original Blandford-Znajek scheme, the field lines are anchored in
the event horizon, but extend to infinity.   It is also possible that some or
all the field lines threading the horizon could instead close by passing through
the disk.  Because the material in the disk will arrive at the
black hole in an infall time, the details of such a structure must be transitory.  However, by the same token, if the loops close well outside the
inner edge of the disk, the characteristic duration of this structure would
be $\sim \alpha^{-1} (r/h)^2$ dynamical times, where $\alpha$ is the usual
ratio of $r$--$\phi$ stress to local pressure and $h$ is the disk thickness.
Particularly in a thin disk, this could be a relatively long time.

   For this reason, some discussions of this scheme have assumed a time-steady
situation and its corollary, field lines with a fixed rotation frequency
(e.g., \cite{jhk:l00a}; but see \cite{jhk:g00} for an intrinsically non-steady
version).  In this case, because the resistance of disk plasma
is so much less than $R_h$, $\Omega_F$ for all field-lines
must be very nearly the rotational frequency $\Omega_d$
of the disk matter they thread.  To estimate the offset, we observe that
the $\vec J \times \vec B$ force can be written in two equivalent ways.
Most often, this force is evaluated by using Amp\`ere's Law (neglecting
the displacement current) to write
$\vec J = (c/4\pi)\nabla \times \vec B$.  However, one can also use
Ohm's Law to compute the current, i.e. set
$\vec J = \sigma(\vec E + \vec v/c \times \vec B)$.  Identifying the field-line
velocity with the $\vec E \times \vec B$ drift speed \cite{jhk:j75}, we can
then equate the two forms and find
\begin{equation}
|\Omega_F - \Omega_d| \sim {c^2 \over 8\pi h r \sigma}
\end{equation}
where $h$ is the disk thickness and $r$ is the radial coordinate of
the place the field-lines pass through.  If resistivity
in accretion disks is due to electron-ion Coulomb collisions,
$\sigma \sim 10^{14}$~s$^{-1}$; even if particle-wave scattering or
turbulence magnifies the effective collision rate, $|\Omega_F - \Omega_d|
\ll \Omega_d$.

     To compute the rate at which black hole rotational energy is tapped,
one must next estimate the total resistive load due to the disk.
The energy per unit volume dissipated by true
electrical resistance is the characteristic rate of dissipation for
turbulent magnetic field in the disk, $\sim (B^2/8\pi) \Omega $, times
a very small factor $\sim c^2/(v_{orb}\sigma h)$.  On the other hand,
if $\langle B_r B_\phi \rangle \neq 0$, the rotational work done by the
field is likely to be much larger.  The problem is that the field
structure within the disk is very much {\it not} force-free, so
the usual simplifications invoked to estimate power output in the
Blandford-Znajek mechanism don't apply.  As a result, detailed calculations
are necessary to evaluate the true load---and therefore the
power that is actually generated---in this model.
 
\section{Field Lines Anchored in Plunging Plasma, Closed in the Disk,
and Not Force-Free}

    In contrast to the previous two pictures, let us now
consider the consequences of a disk whose material is actually
accreting.  Within it, the
growth of the magneto-rotational instability to nonlinear amplitude creates
strong MHD turbulence.  When plasma follows circular orbits whose
frequency declines outward, the shear automatically biases the turbulence
so that $\langle B_r B_\phi \rangle < 0$ and angular
momentum is carried outward,
thereby permitting accretion \cite{jhk:bh98}.

    When material leaves the marginally stable orbit and plunges inward, its
high electrical conductivity assures that it carries magnetic flux along.
Whether or not there is any net magnetic flux threading the black hole's
event horizon, most magnetic field lines in the plunging plasma are closed
relatively close by, for the MHD turbulence in the disk should lead to much
reconnection.  The question now arises as to whether these field lines,
threading plasma plunging along unstable orbits through the ergosphere,
can exert enough stress to tap the black hole's rotational energy at
an interesting rate.

     Novikov \& Thorne \cite{jhk:nt73} argued that in
a time-steady, axi-symmetric disk
the stress $T_{r\phi}$ should approach zero at the marginally stable orbit
because the
inertia of material in the plunging region inside $r_{ms}$ would be too small to
exert any significant stress.  If this were the case, this version of the
Blandford-Znajek mechanism (and the previous one!) would be unimportant. 
However, as remarked in \cite{jhk:t74,jhk:pt74},
it is possible for magnetic forces to
be significant even in the absence of substantial inertia.  If so, $T_{r\phi}$
need not go to zero at $r_{ms}$.

    In fact, the same MHD processes that account for angular momentum transport
in the disk proper are likely to apply in the neighborhood of $r_{ms}$.
The existence of the magneto-rotational instability depends only on the
disk shear; it is therefore just as strong in the vicinity of $r_{ms}$ as
it is farther out in the disk.  There is also no particular reason why MHD
turbulence should damp more quickly in the inner disk than the outer.  We
may therefore expect the amplitude of MHD turbulence (normalized to the
disk pressure) to remain roughly constant as $r \rightarrow r_{ms}$.
Because the shear is still strong, we can likewise expect $T_{r\phi}/Tr(T)$
to not change dramatically in this region.  In fact, if anything, the
ratio of magnetic stress to pressure is likely to rise near $r_{ms}$
because the pressure begins to decline as the inflow velocity
rises, whereas we can expect the magnetic field strength to vary more
slowly.  It therefore appears that there
is no local mechanism to enforce the decline in $T_{r\phi}$ demanded by
the Novikov-Thorne model.

      If this is the case, flux-freezing
through the plunging region can lead to magnetic forces
becoming comparable to gravity in a roughly time-steady and axi-symmetric
accretion flow when $v_r \sim v_\phi$ \cite{jhk:k99}.
$T_{r\phi}$ at $r_{ms}$ large enough to significantly alter the accretion
efficiency follows as a corollary \cite{jhk:g99};
in a time-steady, axi-symmetric state, the additional dissipation
per unit area in the disk falls roughly as $r^{-7/2}$ at $r \ge r_{ms}$
\cite{jhk:ak00}.

    Recent simulations by Hawley \& Krolik \cite{jhk:hk00}
(3-d non-relativistic MHD
in the pseudo-Newtonian Pacy\'nski-Wiita potential) support these arguments
(but see also \cite{jhk:arc00}).
They find that the inner regions of disks become highly turbulent and
non-steady, but that the azimuthally-averaged and vertically-integrated
magnetic stress is essentially flat from the innermost part of the disk
proper through the marginally stable orbit and well into the plunging
region.  One might therefore expect the accretion efficiency to be
somewhat greater than the one predicted on the basis of zero-stress
at $r_{ms}$ even without black hole rotation.

    Black hole rotation (and its attendant frame-dragging) may enhance
this supplemental energy release because it is mediated by magnetic
torques \cite{jhk:g99}.  In other words, when the black hole rotates,
this mechanism is very similar to the previous one, but transfers the inner
field ``anchor" to plunging plasma and does not require either
force-free field structure or time-steadiness.

     As for any mechanism proposed to tap black hole spin, ``spindown"
must be defined carefully.
If the weakest of the three criteria is used,  {\it any}
increase in the accretion efficiency above the nominal is associated with a
diminution of the black hole's rotational energy, for it means that the
accreting matter enters the hole with smaller angular momentum than it might
otherwise have.  If the constant $a_*$ criterion is used and captured
radiation is ignored, spindown occurs when the efficiency is greater than
\begin{equation}
\eta_{eq} = 1 - {\sqrt{ 1 - 3/r_{ms} + 2a_*/r_{ms}^{3/2}} \over
 1  - a_*/r_{ms}^{3/2}},
\end{equation}
\cite{jhk:ak00}.  For $0 \le a_* \ltwid 0.95$,
$\eta_{eq} \simeq 0.3$ -- 0.35.  Captured
radiation diminishes $\eta_{eq}$ by $\simeq 0.1$ for $a_* \gtwid 0.99$.
The contrast between this criterion and the strongest one (in which the
rotational energy actually declines) is illustrated in Figure 1.  Although
a genuine evaluation of how large this effect might be requires numerical
MHD simulations, simplified (simplistic?) analytic estimates indicate
that even this criterion might be met when $a_* \gtwid 0.6$ \cite{jhk:g99}.

\begin{figure}
\centerline{\psfig{file=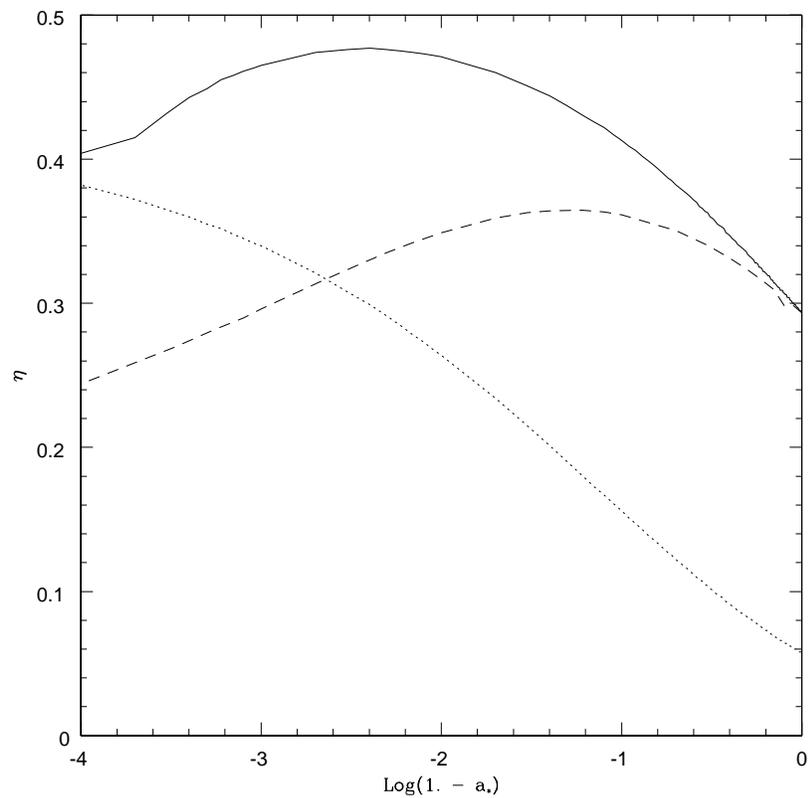, height=4.5in, width=4.5in}}
\caption{
Fig.~1.\  The efficiency $\eta$ above which the rotational energy is diminished
is shown by a solid curve.  For comparison, the dotted curve shows
the Novikov-Thorne efficiency; the dashed curve shows the efficiency at which
$a_*$ does not change \protect\cite{jhk:ak00}.  Both the solid and dashed curves
are specific to the model of this section.}
\end{figure}

    Because the spindown luminosity in this mechanism is intimately tied
to accretion, its natural luminosity scale is the accretion luminosity scale,
$\sim \dot M_o c^2$.  However, effects such as mass clumping or magnetic reconnection could diminish it, possibly by sizable factors.

\section{Field Loops Anchored in the Ergosphere, Closed inside the Marginally Stable Orbit, and Not Force-Free }

    If, like the magnetic field on the horizon, the field strength in the plunging region is comparable to its value in the disk, it could
in principle drive a substantial magneto-centrifugal wind.  Unlike most
MHD wind models, this one would be very far from time-steady, for
the character of the field in this region changes in a free-fall time.
In a time-steady wind, field lines must extend to infinity, for fluid
elements have been travelling outward, carrying their magnetic flux with
them, for effectively forever.  By contrast, when the wind is highly
variable, there is no requirement for any continuity in field structure
between fluid elements expelled long ago and those just entering the wind
today.

Nonetheless, the same dynamics that drive time-steady winds should also
apply in this context.  In fact, circumstances might be more propitious
toward wind creation in this region than in the disk, for
the Alfven speed should be rather higher due to the similar strength
magnetic field and the much smaller inertia (cf. the simulations reported
in \cite{jhk:m99}).  Local fluctuations in the angular momentum of plunging fluid
elements could lead to shocks in this region, and strong local heating.
Just as for the Blandford-Znajek mechanism,
the characteristic luminosity scale is $\sim |\vec B|^2 r_g^2 c$, but the
coefficient could be considerably less than unity.

\section{Disk Winds}

     Finally, it is useful as a standard of comparison to contrast these
variations on the Blandford-Znajek mechanism with magneto-centrifugal winds
launched from an accretion disk.  Indeed, Livio et al. \cite{jhk:lop98}
claimed that the black hole spin energy reservoir would {\it always} be
relatively unimportant because electromagnetic
energy loss from the accretion disk would always exceed that from the black
hole itself through the Blandford-Znajek mechanism.  Their argument
(an elaboration of one in \cite{jhk:bz77}) hinged
on a simple order-of-magnitude comparison: They estimated that
the luminosity of a disk-driven wind would be generically
$\sim O(10) c r_g^2 B_{pd} B_{td}$, where $B_{(p,t)d}$ are the poloidal and toroidal
magnetic field components in the disk and the factor $O(10)$ accounts for
the fact that the surface area of the disk is rather larger than the surface
area of the black hole.  Supposing that all currents supporting magnetic
fields are found in the disk, they then estimated $B_h \ltwid B_{pd}$, so
that the disk luminosity would always be greater than the Blandford-Znajek
luminosity.

    However, this luminosity estimate is based on the implicit assumption that
the $r$-$\phi$ component of the Maxwell stress in the disk wind has a vertical
scale height comparable to the disk radius,
not the (gas density-determined) disk thickness.  This is
a very strong assumption.  At the moment, there is no way to rigorously
evaluate its quality, but there are a few suggestive results in hand
already from simulations \cite{jhk:ms00,jhk:hk00}.
Both of these indicate that the magnetic stresses decline vertically more
slowly than does the gas density, but the scale height ratio is only a factor
of two or three.  If later work confirms these tentative results, the estimated
magnetic wind luminosity of the disk would be greatly reduced.
Mechanisms that derive their energy from the spin of the black hole, but
do not put the energy into winds, would then rise in relative importance.

\section{Conclusions}

      Virtually every idea anyone has suggested about how to extract spin energy
from black holes involves magnetic fields anchored in the black hole's
ergosphere in one way or another.  At the same time, the variations possible
among these ideas (whether the field lines close at infinity or in the disk,
whether they are anchored in the horizon or farther away, $\ldots$) raise
different questions and different opportunities.  In discussions of these
ideas, it is therefore important to distinguish carefully between them, for
questions and criticisms raised about one version do not necessarily apply
to the others.  It has been the aim of this short paper to provide a convenient
framework for making these distinctions, and to assess (in a necessarily
subjective fashion) their current standing.

    In order for further progress to be made, it is apparent that (at least)
three key questions must be answered:

\noindent 1.) Is it possible for energy and angular momentum to be carried away
from a black hole when the field lines are anchored only in the horizon?

\noindent 2.) Do black holes accumulate net magnetic flux (so that field
loops close far away from the black hole and the structure changes on
very long timescales) or does enough reconnection accompany accretion that
field loops all close nearby, and the structure changes on the infall
timescale?

\noindent 3.) Can one think of the field as being force-free everywhere but
in a small load, or is the plasma inertia qualitatively important?

\acknowledgments

     I am happy to acknowledge stimulating and instructive conversations
with Eric Agol, Roger Blandford, Doug Eardley, and Ethan Vishniac.
This work was partially supported by NSF Grant AST-9616922.

\end{document}